\documentclass[conference,10pt]{IEEEtran}
\usepackage{epsfig,rotating,setspace,latexsym,amsmath,epsf,amssymb,amsfonts,bm,theorem,subfigure,epstopdf,cite,bbm, authblk} 
\usepackage[ruled,vlined]{algorithm2e}

\usepackage{color}

\SetArgSty{textnormal}

\IEEEoverridecommandlockouts
\allowdisplaybreaks

\begin{document}
\bstctlcite{IEEEexample:BSTcontrol} 

\title{Sensing-Throughput Tradeoffs with Generative Adversarial Networks for NextG Spectrum Sharing}


\author[1]{Yi Shi}
\author[2]{Yalin E. Sagduyu}
\affil[1]{\normalsize Commonwealth Cyber Initiative, Virginia Tech, Arlington, VA, USA}
\affil[2]{\normalsize National Security Institute, Virginia Tech, Arlington, VA, USA}

\maketitle

\begin{abstract}
Spectrum coexistence is essential for next generation (NextG) systems to share the spectrum with incumbent (primary) users and meet the growing demand for bandwidth. One example is the 3.5 GHz Citizens Broadband Radio Service (CBRS) band, where the 5G and beyond communication systems need to sense the spectrum and then access the channel in an opportunistic manner when the incumbent  user (e.g., radar) is not transmitting. To that end, a high-fidelity classifier based on a deep neural network is needed for low misdetection (to protect incumbent users) and low false alarm (to achieve high throughput for NextG). In a dynamic wireless environment, the classifier can only be used for a limited period of time, i.e., coherence time. A portion of this period is used for learning to collect sensing results and train a classifier, and the rest is used for transmissions. In spectrum sharing systems, there is a well-known tradeoff between the sensing time and the transmission time. While increasing the sensing time can increase the spectrum sensing accuracy, there is less time left for data transmissions. In this paper, we present a generative adversarial network (GAN) approach to generate synthetic sensing results to augment the training data for the deep learning classifier so that the sensing time can be reduced (and thus the transmission time can be increased) while keeping high accuracy of the classifier. We consider both additive white Gaussian noise (AWGN) and Rayleigh channels, and show that this GAN-based approach can significantly improve both the protection of the high-priority user and the throughput of the NextG user (more in Rayleigh channels than AWGN channels). 
\end{abstract}

\section{Introduction}\label{sec:Introduction}


To improve spectrum utilization, spectrum co-existence has been widely accepted in designing new systems that need to share the spectrum with incumbent users.
For example, in the 3.5 GHz Citizens Broadband Radio Service (CBRS) band, three types of users, incumbent users, Priority Access License (PAL) users, and General Authorized Access (GAA) users, share the spectrum \cite{CBRS}. 
Once the Environmental Sensing Capability (ESC) detects the incumbent user (such as radar), the Spectrum Access System (SAS) is informed to reconfigure the access to the spectrum. As the spectrum environment is complex, deep learning provides a powerful capability to detect wireless signals of interest, whereas conventional machine learning algorithms cannot typically achieve high accuracy especially in the low signal-to-noise-ratio (SNR) regime \cite{DL1, DL2, DL3, DL4, DL5}. 

In this spectrum sharing system, the NextG user needs to sense the incumbent user activities and transmit only if the incumbent user is detected. 
We assume that time is slotted and incumbent user will not change its busy/idle status within a time slot. Then, the NextG user can sense channel at the beginning of a slot and apply a classifier to detect incumbent user activities.
In this paper, we consider a deep neural network as the deep learning-based classifier. The NextG user starts with a learning phase to collect sensing results and train a deep neural network  as the classifier that is then used in the subsequent transmission phase.

If the incumbent user channel is static, the trained classifier can be always used without loss on accuracy.
However, the incumbent user channel may change over time (e.g., due to the mobility) and thus the accuracy of the trained classifier may decrease.
To achieve good throughput and to protect the incumbent user transmissions, the NextG user needs to retrain its classifier over time.
We assume that the NextG user needs to retrain its classifier periodically to accommodate potential changes in channel, traffic, and interference effects, i.e., the NextG user trains a classifier in the learning phase at the beginning of each time frame and then uses this classifier in the transmission phase (the rest of a time frame). The length of a time frame is determined by the incumbent user channel coherent time.
It is clear that a long sensing phase increases the accuracy of the incumbent user detection (by collecting more sensing samples), but the remaining time for the NextG user transmission becomes limited. On the other hand, a short sensing phase may not be sufficient to train a classifier with high accuracy. Thus, there is a well-known fundamental trade-off between sensing and transmission phases \cite{Liang08:sensing}.

In this paper, we improve the balance of this sensing-throughput tradeoff by applying a generative adversarial network (GAN) to generate synthetic sensing results and use both real and synthetic sensing results as the input to train a deep learning-based classifier.
The GAN was used originally introduced for computer vision applications \cite{goodfellow} and then later extended to generate synthetic data in the wireless domain for various purposes.
The GAN was used to augment training data to improve wireless signal classifiers \cite{Davaslioglu18:GAN} and modulation classifiers \cite{Patel20:GAN}.
In \cite{Roy19:GAN}, the GAN was used to generate synthetic signals for various RF environments.
In \cite{Shi21:GAN, Shi19:GAN}, the GAN was used to generate synthetic signals for spoofing attacks. 
In \cite{Ayanoglu22:GAN}, the GAN was used to train a wireless anomaly/outlier detector.
In \cite{Erpek19:GAN}, the GAN was used to assist a jammer to determine whether there is a transmission to be jammed. In \cite{AntiJamming:GAN}, the GAN was used to enhance an anti-jamming spectrum access scheme. The GAN was also used to generate wireless signals for camouflage \cite{Hou21:GAN} or against user equipment (UE) identification \cite{Sagduyu21:GAN}. 
In \cite{Davaslioglu22:GAN}, the GAN was used to train an optimizer an autoencoder communications system for interference suppression.
In \cite{Oshea19:model} and \cite{Yang19:data}, the GAN was used to model and generate wireless channel data.
In \cite{Ye20:CGAN}, the GAN was used to generate channel status information.

In this paper, we use the GAN to generate more sensing results to train a high-fidelity detector for the incumbent user activities. 
In \cite{Hu20:train}, the GAN was used to generate more channel data samples to improve channel estimation with limited sensing data.
In \cite{Tang18:GAN}, the GAN was used to generate more data samples to train a modulation classifier.
Compared to the previous studies that have focused on the data generation objectives, our goal in this paper is to study the effect of the GAN on the sensing-throughput trade-offs by shortening the length of the learning phase via the GAN and thus obtaining a long transmission phase for the NextG user to achieve high throughput.
Our results show that the GAN can improve the incumbent user detection (in terms of small misdetection probability) and the NextG user throughput compared with the scheme without the GAN for both AWGN and Rayleigh channels. Our results indicate that the GAN can be effectively used to improve spectrum utilization for NextG spectrum sharing systems without extra communication or sensing capabilities. 

The remaining of this paper is organized as follows.
Section~\ref{sec:model} presents the NextG spectrum coexistence system, where the sensing process may be enhanced by the GAN to improve the throughput.
Section~\ref{sec:result} evaluates the performance with and without the GAN.
Section~\ref{sec:conclude} concludes this paper.

\section{System Model}
\label{sec:model}

\begin{figure}
	\centering
	\includegraphics[width=1\columnwidth]{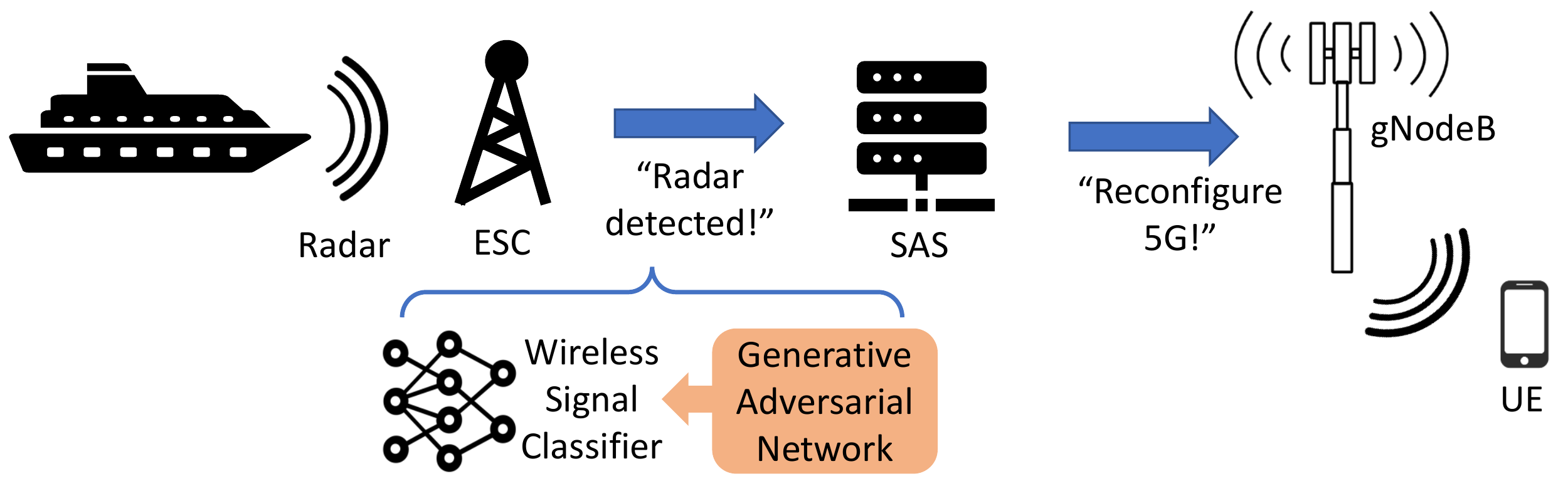}
	\caption{Example of NextG spectrum sharing.}
	\label{fig:5G}
\end{figure}

We consider a NextG spectrum sharing scenario (illustrated in Fig.~\ref{fig:5G}), where the NextG user applies a deep learning-based classifier (namely, a deep neural network) to detect the ongoing incumbent user transmissions. This classifier needs to be retrained periodically due to changes in the incumbent user signals. We consider the coherence time, namely the time period when the channel characteristics remains the same, as a time frame. The NextG user needs to  train a classifier first and then uses this classifier in the remaining of the time frame. The problem is how to determine a suitable training period so that (i) the achieved throughput is maximized and (ii) the impact on the incumbent user is minimized. If the training period is too short, the classifier may have a high false alarm probability (the transmission opportunity is lost) and a high misdetection probability (the impact on the incumbent user is large). If the training period is too long, the remaining time is short and thus  the NextG user may not achieve a large throughput. The optimal length of the training period has been studied for cognitive radio networks \cite{Liang08:sensing}. 

In this paper, we utilize the GAN to generate additional sensing results and then use both real and synthetic results to train a classifier. We will show that with the GAN, the training period can be made shorter and the system performance can be improved. Table~\ref{table:notation} shows the notation used in this paper.

\begin{table}
	\caption{Notation.}
	\centering
	{\small
		\begin{tabular}{c|l} 
$T$ & The length of a time frame (coherence time for a \\
    & trained classifier) \\ \hline
$S$ & The length of a time slot  \\ \hline
$F$ & The number of sensing results in a sample \\ \hline
$L$ & The number of slots in the learning period \\ \hline
$\hat L$ & The reduced number of slots in the learning period by \\
         & using GAN \\ \hline
$p_s$ & The fraction of time in a slot to collect sensing results \\
      & in the training period \\ \hline
$r$ & Throughput \\ \hline
$\hat r$ & Throughput by using GAN \\ \hline
$p_B$ & The probability that incumbent user is busy \\ \hline
$p_{FA}$ & The false alarm probability \\ \hline
$\hat p_{FA}$ & The false alarm probability by using GAN \\ \hline
$p_N$ & The failure probability due to random large noise \\ \hline
$p_{MD}$ & The misdetection probability \\ \hline
$\hat p_{MD}$ & The misdetection probability by using GAN \\ \hline
$n$ & The number of synthetic samples
		\end{tabular}
	}
	\label{table:notation}
\end{table}

\subsection{Sensing and Transmission Scheme}

\begin{figure}
	\centering
	\includegraphics[width=1\columnwidth]{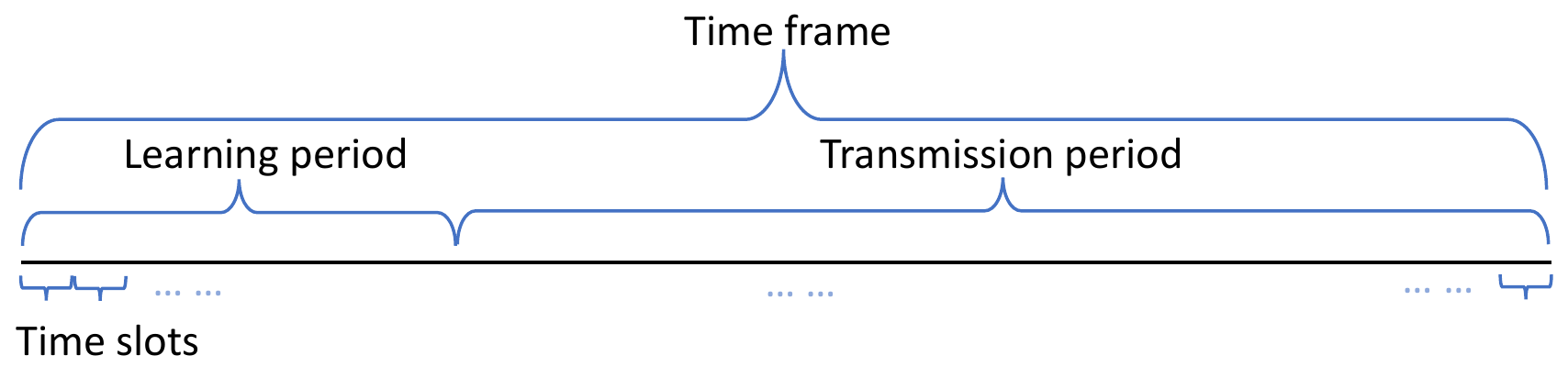}
	\caption{The time structure.}
	\label{fig:time}
\end{figure}

The time structure is shown in Fig.~\ref{fig:time}. The length of a time frame is $T$, where we normalize the time for obtaining one sensing result as $1$. In this time frame, there are many time slots and the incumbent user does not change its busy/idle status within a time slot. Denote the time slot length as $S$. 
\begin{itemize}
    \item The training time uses $L$ time slots to collect data (namely, power levels of received signals) for deep learning. The NextG receiver senses most of the time (e.g., $p_s=85$\%) in a slot to collect sensing data. 
   To determine the busy/idle status, the NextG transmitter transmits in the remaining time. If successfully received (for most of transmissions), this slot is considered as idle, otherwise this slot is considered as busy. 
   At the end of the learning period, the NextG receiver has many sensing results for idle/busy slots. The results in the same class are grouped with size $F$ to construct machine learning samples. With samples and their class (status) information, the NextG receiver can build a deep learning-based classifier to predict the channel status.
    
	\item The remaining of the frame is used for transmissions. Within a slot, the NextG receiver first senses $F$ power levels, then runs the classifier to determine the channel status. If the channel is idle, the NextG transmitter transmits data in the remainder of this slot.
\end{itemize}

Assume the throughput by successful transmission over $1$ unit of time is $1$. Then, the throughput can be calculated as
\begin{equation}
r=(T-LS)(1-p_B)(1-p_{FA})\left(\frac{S-F}{S}\right) \left(1-p_N\right) \; ,  
\end{equation}
where $p_B$ (e.g., $0.5$) is the probability that the incumbent user is busy, $p_{FA}$ is the false alarm probability, and $p_N$ is the failure probability due to random large noise (e.g., $5$ as a threshold for noise with mean $0$ and variance $1$ and incumbent user signal with mean $3$ and variance $1$). We need to determine $L$, which in turn will determine the transmission time $T-LS$ and the throughput $r$. Another performance metric is the misdetection probability $p_{MD}$ (or the protection of the incumbent user), which also depends on $L$.
A large $L$ can reduce $p_{MD}$ but also reduce $T-LS$.
Thus, there is a tradeoff between the NextG user throughput and the incumbent user protection when we determine $L$.

\subsection{GAN-Enhanced Spectrum Sensing Scheme}

\begin{figure}
	\centering
	\includegraphics[width=1\columnwidth]{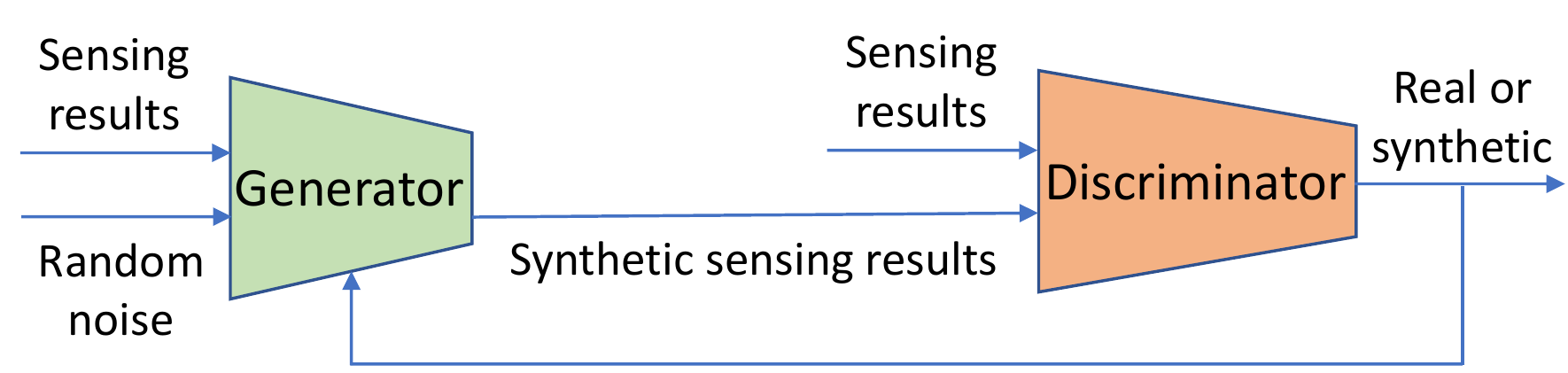}
	\caption{The GAN structure.}
	\label{fig:GAN}
\end{figure}

Instead of just considering a tradeoff when we determine $L$, we can employ the GAN to achieve better performance.
The GAN is trained at the end of the learning period. We then use the GAN to generate more samples to train a classifier.
The GAN consists of a deep neural network as the generator and a deep neural network as the discriminator (see Fig.~\ref{fig:GAN}). The generator can generate synthetic sensing results (statistically similar to real sensing results) by using real sensing results and some random noise as input.
The discriminator classifies sensing results as real or synthetic. 
During the training of the GAN, the classification results are provided to the generator as feedback.
Once trained, the generator can generate synthetic sensing results that cannot be distinguished from real. 

Since the GAN is used to generate more samples at the end of training period and the NextG user uses both real and synthetic samples as training data to build a deep learning-based classifier, it is expected that this classifier can achieve better performance in terms of false alarm probability $\hat p_{FA}$ and misdetection probability $\hat p_{MD}$ while using less number of learning slots $\hat L$. Then, the throughput is updated as 
\begin{equation}
    \hat r = (T- \hat L S)(1-p_B)(1- \hat p_{FA})\left(\frac{S-F}{S}\right)(1-p_N) \; .
\end{equation}
For easy comparison, the generated samples will be the same as the samples that the NextG receiver can obtain by using some additional slots. That is, suppose that the NextG user uses $n$ additional slots and obtains around $\frac{nSp_s}{F}$ samples. To be comparable, the number of generated samples is $\lfloor \frac{nSp_s}{F} \rfloor$ and we say that these synthetic samples are generated for $n$ synthetic slots.
The time for training a GAN is similar to the time for training a classifier. 

\section{Performance Evaluation}
\label{sec:result}
We consider a time frame of $T=20,000$ and a slot has length $S=20$. During the learning period, $p_s=85$\% of a slot is used to collect sensing results and the remaining time is used to determine the channel status (the incumbent user's busy/idle status).
After the learning period, $F=5$ sensing results are collected at the beginning of each slot to determine the channel status. The probability that the incumbent user is active is $p_B=0.5$. We consider Gaussian and Rayleigh channels. We use a feedforward neural network (see Table~\ref{table:classifier}) to build the classifier. After hyperparameter optimization, we end up with three hidden layers with $128, 64$, and $32$ neurons. The ReLU activation function is used at each hidden layer. After each hidden layer, we add a dropout layer with rate $0.2$ to avoid overfitting. The last layer has two neurons (corresponding to two classes) and uses a Softmax activation function. We use categorical cross-entropy as the loss function, Root Mean Squared Propagation (RMSprop) as the optimizer, and accuracy as the optimization metric.

\begin{table}
	\caption{The deep neural network structure of the classifier.}
	\centering
	{\small
		\begin{tabular}{c|c|c|c}
		layer & number of & activation & dropout \\
		 & neurons & function & rate \\ \hline \hline
Hidden 1 & 128 & Relu & 0.2 \\ \hline
Hidden 2 & 64 & Relu & 0.2 \\ \hline
Hidden 3 & 32 & Relu & 0.2 \\ \hline
Output & 2 & Softmax & 
		\end{tabular}
	}
	\label{table:classifier}
\end{table}

The GAN model to generate synthetic sensing results consists of two deep neural networks, one for the generator and one for discriminator.
The structure of these two networks is shown in Table~\ref{table:gan}.
The generator has three hidden layers, each with $128$ neurons. The output layer has $S p_s$ neurons, i.e., its output constitutes the synthetic sensing results in a slot.
The discriminator also has three hidden layers, each with $128$ neurons. The output layer has one neuron with Sigmoid activation function, i.e., its output is the probability that the input sample is real.
The loss function is sigmoid cross-entropy, which is used as the optimization metric.
We use Adam as the optimizer.
Once trained, the generator provides synthetic samples as the output. Then, we build a classifier using the same feedforward neural network architecture that we used to build a classifier without the GAN.

\begin{table}
	\caption{The GAN neural network structure.}
	\centering
	{\small
		\begin{tabular}{c|c||c|c}
		\multicolumn{2}{c||}{Generator} & \multicolumn{2}{c}{Discriminator} \\ \hline
		parameter & value & parameter & value \\ \hline
Hidden & 3 $\times$ 128 & Hidden & 3 $\times 128$ \\ \hline
Output & $S p_s$ & Output & 1 \\ \hline
Activation & Sigmoid & Activation & Sigmoid \\ \hline
Loss & crossentropy & Loss & crossentropy \\ \hline
Optimizer & Adam & Optimizer & Adam
		\end{tabular}
	}
	\label{table:gan}
\end{table}


\subsection{Results for AWGN Channel}

We start with Gaussian channel. In the first set of results, we fix the training time $\hat L=10$ and consider different number of GAN training data samples so that the total training time is varied from $L=15$ to $50$. Table~\ref{table:l10} shows the comparison with and without the GAN. The use of the GAN for sensing data augmentation 
may
improve the throughput as well as the misdetection probability. We also observe that 
the classifier built upon too many synthetic samples may not achieve high accuracy. As a result, both the throughput and the misdetection probability performance may not be satisfactory.
By considering both throughput and misdetection probability objectives, the best choice is to have $n=15$ synthetic samples generated by the GAN.

\begin{table}
	\caption{Comparison of the performance with and without the GAN for AWGN channel and $\hat L=10$.}
	\centering
	{\small
		\begin{tabular}{c|c|c||c|c|c}
		\multicolumn{3}{c||}{No GAN} & \multicolumn{3}{c}{GAN} \\ \hline
		$L$ & $r$ & $p_{MD}$ (\%) & $\hat L+n$ & $\hat r$ & $\hat p_{MD}$ (\%) \\ \hline
15 & 6912 & 17.83 & 10+5 & 7252 & 23.33 \\ \hline
20 & 7016 & 15.81 & 10+10 & 7180 & 13.44 \\ \hline
25 & 6928 & 13.61 & 10+15 & 7208 & 11.20 \\ \hline
30 & 7032 & 8.73 & 10+20 & 7134 & 22.00 \\ \hline
35 & 7004 & 10.00 & 10+25 & 7034 & 6.11 \\ \hline
40 & 6932 & 11.46 & 10+30 & 7005 & 13.24 \\ \hline
45 & 6887 & 7.13 & 10+35 & 7022 & 16.70 \\ \hline
50 & 6814 & 10.92 & 10+40 & 7064 & 12.22
		\end{tabular}
	}
	\label{table:l10}
\end{table}

Next, we vary $\hat L$. For $\hat L=15$, we show the results with and without the GAN in Table~\ref{table:l15}. Again, using the GAN can improve the throughput as well as the misdetection probability. By considering both throughput and misdetection probability measures, the best choice is to have $n=35$ synthetic samples. For $\hat L=20$, we show the results with and without the GAN in Table~\ref{table:l20}.
By considering both throughput and misdetection probability measures, the best choice is to have $n=80$ synthetic samples. For $\hat L=25$, we show the results with and without the GAN in Table~\ref{table:l25}.
By considering both throughput and misdetection probability measures, the best choice to have $n=115$ synthetic samples. For $\hat L=30$, we show the results with and without the GAN in Table~\ref{table:l30}.
By considering both throughput and misdetection probability measures, the best choice is to have $n=130$ synthetic samples.

\begin{table}
	\caption{Comparison of the performance with and without the GAN for AWGN channel and $\hat L=15$.}
	\centering
	{\small
		\begin{tabular}{c|c|c||c|c|c}
		\multicolumn{3}{c||}{No GAN} & \multicolumn{3}{c}{GAN} \\ \hline
		$L$ & $r$ & $p_{MD}$ (\%) & $\hat L+n$ & $\hat r$ & $\hat p_{MD}$ (\%) \\ \hline
20 & 7016 & 15.81 & 15+5 & 7180 & 13.44 \\ \hline
30 & 7032 & 8.73 & 15+15 & 7134 & 22.00 \\ \hline
40 & 6932 & 11.46 & 15+25 & 7005 & 13.24 \\ \hline
50 & 6814 & 10.92 & 15+35 & 7064 & 12.22 \\ \hline
60 & 6756 & 10.17 & 15+45 & 7193 & 21.31 \\ \hline
70 & 6697 & 7.07 & 15+55 & 7251 & 21.11 \\ \hline
80 & 6625 & 8.87 & 15+65 & 7192 & 10.45 \\ \hline
90 & 6581 & 5.91 & 15+75 & 7163 & 11.27 \\ \hline
100 & 6508 & 13.05 & 15+85 & 7236 & 18.24
		\end{tabular}
	}
	\label{table:l15}
\end{table}

\begin{table}
	\caption{Comparison of the performance with and without the GAN for AWGN channel and $\hat L=20$.}
	\centering
	{\small
		\begin{tabular}{c|c|c||c|c|c}
		\multicolumn{3}{c||}{No GAN} & \multicolumn{3}{c}{GAN} \\ \hline
		$L$ & $r$ & $p_{MD}$ (\%) & $\hat L+n$ & $\hat r$ & $\hat p_{MD}$ (\%) \\ \hline
40 & 6932 & 11.46 & 20+20 & 7061 & 13.55 \\ \hline
60 & 6756 & 10.17 & 20+40 & 7090 & 13.14 \\ \hline
80 & 6625 & 8.87 & 20+60 & 7193 & 11.91 \\ \hline
100 & 6508 & 13.05 & 20+80 & 7149 & 9.86 \\ \hline
120 & 6405 & 15.87 & 20+100 & 7119 & 14.17 \\ \hline
140 & 6288 & 13.95 & 20+120 & 7134 & 8.83 \\ \hline
160 & 6142 & 10.29 & 20+140 & 7150 & 12.11 \\ \hline
180 & 5994 & 8.31 & 20+160 & 7134 & 9.24 \\ \hline
200 & 5831 & 11.03 & 20+180 & 7119 & 7.19
		\end{tabular}
	}
	\label{table:l20}
\end{table}

\begin{table}
	\caption{Comparison of the performance with and without the GAN for AWGN channel and $\hat L=25$.}
	\centering
	{\small
		\begin{tabular}{c|c|c||c|c|c}
		\multicolumn{3}{c||}{No GAN} & \multicolumn{3}{c}{GAN} \\ \hline
		$L$ & $r$ & $p_{MD}$ (\%) & $\hat L+n$ & $\hat r$ & $\hat p_{MD}$ (\%) \\ \hline
40 & 6932 & 11.46 & 25+15 & 7061 & 14.02 \\ \hline
60 & 6756 & 10.17 & 25+35 & 7061 & 8.45 \\ \hline
80 & 6625 & 8.87 & 25+55 & 6986 & 11.55 \\ \hline
100 & 6508 & 13.05 & 25+75 & 7032 & 10.72 \\ \hline
120 & 6405 & 15.87 & 25+95 & 7075 & 8.45 \\ \hline
140 & 6288 & 13.95 & 25+115 & 7075 & 6.19 \\ \hline
160 & 6142 & 10.29 & 25+135 & 7032 & 6.80 \\ \hline
180 & 5994 & 8.31 & 25+155 & 6986 & 10.10 \\ \hline
200 & 5831 & 11.03 & 25+175 & 7047 & 16.70
		\end{tabular}
	}
	\label{table:l25}
\end{table}

\begin{table}
	\caption{Comparison of the performance with and without the GAN for AWGN channel and $\hat L=30$}
	\centering
	{\small
		\begin{tabular}{c|c|c||c|c|c}
		\multicolumn{3}{c||}{No GAN} & \multicolumn{3}{c}{GAN} \\ \hline
		$L$ & $r$ & $p_{MD}$ (\%) & $\hat L+n$ & $\hat r$ & $\hat p_{MD}$ (\%) \\ \hline
40 & 6932 & 11.46 & 30+10 & 7077 & 6.65 \\ \hline
60 & 6756 & 10.17 & 30+30 & 7077 & 6.24 \\ \hline
80 & 6625 & 8.87 & 30+50 & 6988 & 12.47 \\ \hline
100 & 6508 & 13.05 & 30+70 & 7032 & 6.24 \\ \hline
120 & 6405 & 15.87 & 30+90 & 7017 & 8.32 \\ \hline
140 & 6288 & 13.95 & 30+110 & 7032 & 7.48 \\ \hline
160 & 6142 & 10.29 & 30+130 & 7090 & 3.95 \\ \hline
180 & 5994 & 8.31 & 30+150 & 7060 & 5.61 \\ \hline
200 & 5831 & 11.03 & 30+170 & 7047 & 9.36
		\end{tabular}
	}
	\label{table:l30}
\end{table}

\begin{figure}
	\centering
	\includegraphics[width=0.9\columnwidth]{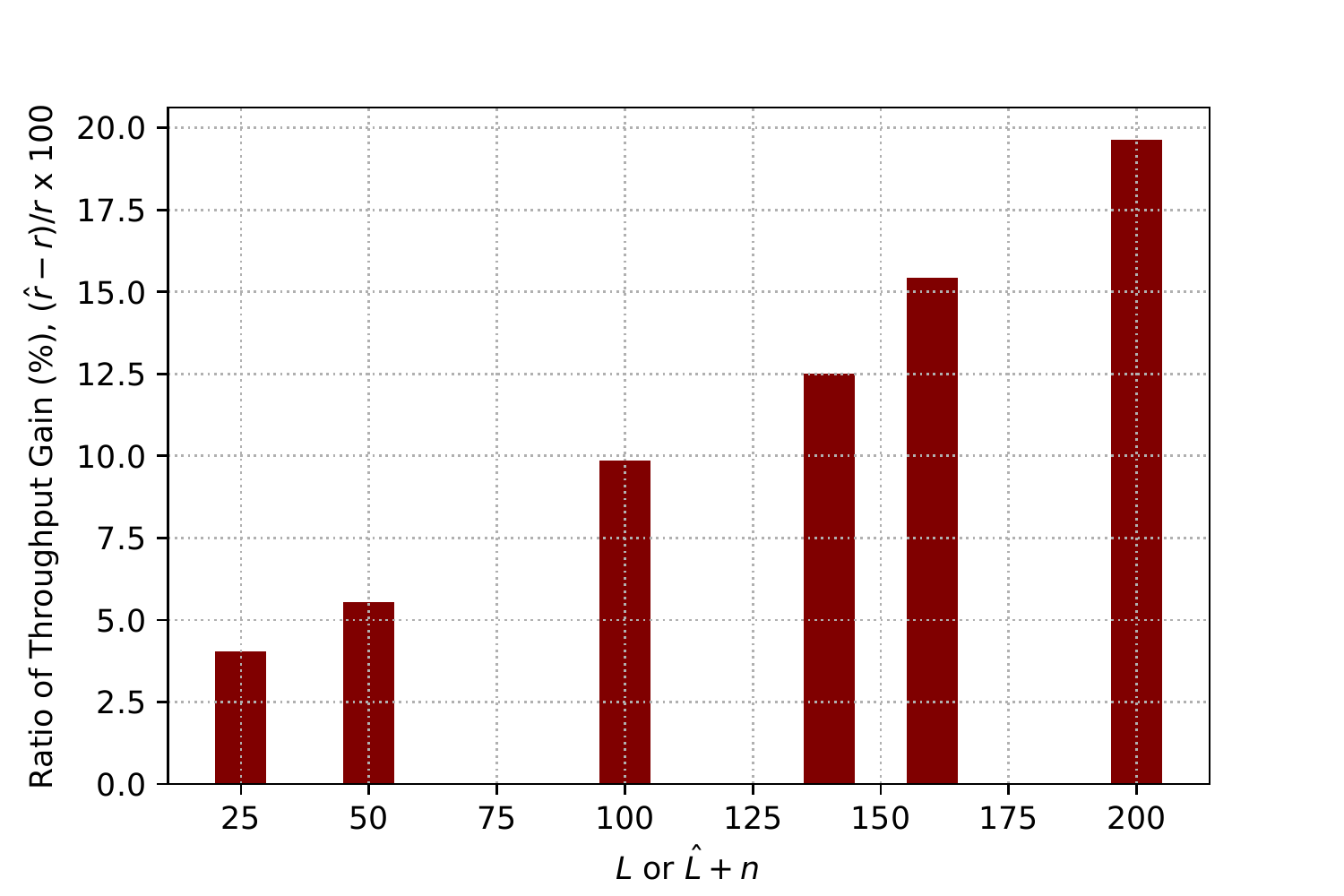}
	\caption{The ratio of throughout improvement for AWGN channel.}
	\label{fig:sensing1}
\end{figure}

We summarize the results for different $L$ (without the GAN) or $\hat L+n$ (with the GAN) in Table~\ref{table:summary}.
Figure~\ref{fig:sensing1} shows the throughput improvement (in terms of $(\hat r -r)/r \times 100$) with and without the GAN. This improvement increases with $L$ (or $\hat L+n$).
With larger $L$ or $\hat L+n$, the throughput is always decreasing. The misdetection probability tends to decrease also.
If $\hat p_{MD} < 10$\%, the choice of $\hat L=15$ and $n=35$ achieves the best results.
If $\hat p_{MD} < 5$\%, the choice of $\hat L=30$ and $n=130$ achieves the best results.

\begin{table}
	\caption{Comparison of the performance with and without the GAN for AWGN channel under different $\hat L$ and $n$.}
	\centering
	{\small
		\begin{tabular}{c|c|c||c|c|c}
		$L$ or & \multicolumn{2}{c||}{No GAN} & \multicolumn{3}{c}{GAN} \\ \cline{2-6}
		$\hat L+n$ & $r$ & $p_{MD}$ (\%) & $n$ & $\hat r$ & $\hat p_{MD}$ (\%) \\ \hline
25 & 6928 & 13.61 & 15 & 7208 & 11.20 \\ \hline
50 & 6814 & 10.92 & 35 & 7192 & 7.17 \\ \hline
100 & 6508 & 13.05 & 80 & 7149 & 9.86 \\ \hline
140 & 6288 & 13.95 & 115 & 7075 & 6.19 \\ \hline
160 & 6142 & 10.29 & 130 & 7090 & 3.95 \\ \hline
200 & 5831 & 11.03 & 160 & 6976 & 7.08
		\end{tabular}
	}
	\label{table:summary}
\end{table}

\subsection{Results for Rayleigh Channel}

Next, we consider Rayleigh channel. The spectrum sensing task for Rayleigh channel is more challenging and thus more time is needed for learning (i.e., larger $\hat L$)  to ensure good performance. We start with $\hat L=50$. Table~\ref{table:l50} shows that the use of the GAN can improve the misdetection probability and the throughput is maximized when $n=100$. However, the misdetection probability is high ($24.59$\%). For $\hat L=100$, Table~\ref{table:l100} shows the results with and without the GAN. By considering both throughput and misdetection probability measures, the best choice is to have $n=200$ synthetic samples. Again, the misdetection probability is high ($17.32$\%). 
For $\hat L=200$, Table~\ref{table:l200} shows the results with and without the GAN. By considering both throughput and misdetection probability measures, the best choice is to have $n=200$ synthetic samples. 
For $\hat L=300$, Table~\ref{table:l300} shows the results with and without the GAN. By considering both throughput and misdetection probability measures, the best choice is to have $n=200$ synthetic samples.

\begin{table}
	\caption{Comparison of the performance with and without the GAN for Rayleigh channel and $\hat L=50$}
	\centering
	{\small
		\begin{tabular}{c|c|c||c|c|c}
		\multicolumn{3}{c||}{No GAN} & \multicolumn{3}{c}{GAN} \\ \hline
		$L$ & $r$ & $p_{MD}$ (\%) & $\hat L+n$ & $\hat r$ & $\hat p_{MD}$ (\%) \\ \hline
100 & 4576 & 35.50 & 50+50 & 3771 & 22.73 \\ \hline
150 & 4381 & 32.79 & 50+100 & 4548 & 24.59 \\ \hline
200 & 4119 & 26.17 & 50+150 & 3987 & 23.76 \\ \hline
250 & 3895 & 26.46 & 50+200 & 4187 & 26.86 
		\end{tabular}
	}
	\label{table:l50}
\end{table}

\begin{table}
	\caption{Comparison of the performance with and without the GAN for Rayleigh channel and $\hat L=100$}
	\centering
	{\small
		\begin{tabular}{c|c|c||c|c|c}
		\multicolumn{3}{c||}{No GAN} & \multicolumn{3}{c}{GAN} \\ \hline
		$L$ & $r$ & $p_{MD}$ (\%) & $\hat L+n$ & $\hat r$ & $\hat p_{MD}$ (\%) \\ \hline
150 & 4381 & 32.79 & 100+50 & 4512 & 28.14 \\ \hline
200 & 4119 & 26.17 & 100+100 & 4454 & 25.32 \\ \hline
300 & 3646 & 20.40 & 100+200 & 4050 & 17.32 \\ \hline
400 & 2956 & 7.82 & 100+300 & 4392 & 23.59 \\ \hline
500 & 2448 & 7.09 & 100+400 & 4432 & 23.81 \\ \hline
600 & 2020 & 9.80 & 100+500 & 4218 & 10.82 
		\end{tabular}
	}
	\label{table:l100}
\end{table}

\begin{table}
	\caption{Comparison of the performance with and without the GAN for Rayleigh channel and $\hat L=200$}
	\centering
	{\small
		\begin{tabular}{c|c|c||c|c|c}
		\multicolumn{3}{c||}{No GAN} & \multicolumn{3}{c}{GAN} \\ \hline
		$L$ & $r$ & $p_{MD}$ (\%) & $\hat L+n$ & $\hat r$ & $\hat p_{MD}$ (\%) \\ \hline
250 & 3895 & 26.46 & 200+50 & 3964 & 13.09 \\ \hline
300 & 3646 & 20.40 & 200+100 & 3916 & 10.86 \\ \hline
400 & 2956 & 7.82 & 200+200 & 3877 & 8.64 \\ \hline
500 & 2448 & 7.09 & 200+300 & 3879 & 9.88 \\ \hline
600 & 2020 & 9.80 & 200+400 & 3866 & 9.63 
		\end{tabular}
	}
	\label{table:l200}
\end{table}

\begin{table}
	\caption{Comparison of the performance with and without the GAN for Rayleigh channel and $\hat L=300$}
	\centering
	{\small
		\begin{tabular}{c|c|c||c|c|c}
		\multicolumn{3}{c||}{No GAN} & \multicolumn{3}{c}{GAN} \\ \hline
		$L$ & $r$ & $p_{MD}$ (\%) & $\hat L+n$ & $\hat r$ & $\hat p_{MD}$ (\%) \\ \hline
400 & 2956 & 7.82 & 300+100 & 3646 & 9.92 \\ \hline
500 & 2448 & 7.09 & 300+200 & 3596 & 9.07 \\ \hline
600 & 2020 & 9.80 & 300+300 & 3478 & 9.63 
		\end{tabular}
	}
	\label{table:l300}
\end{table}

\begin{figure}
	\centering
	\includegraphics[width=0.9\columnwidth]{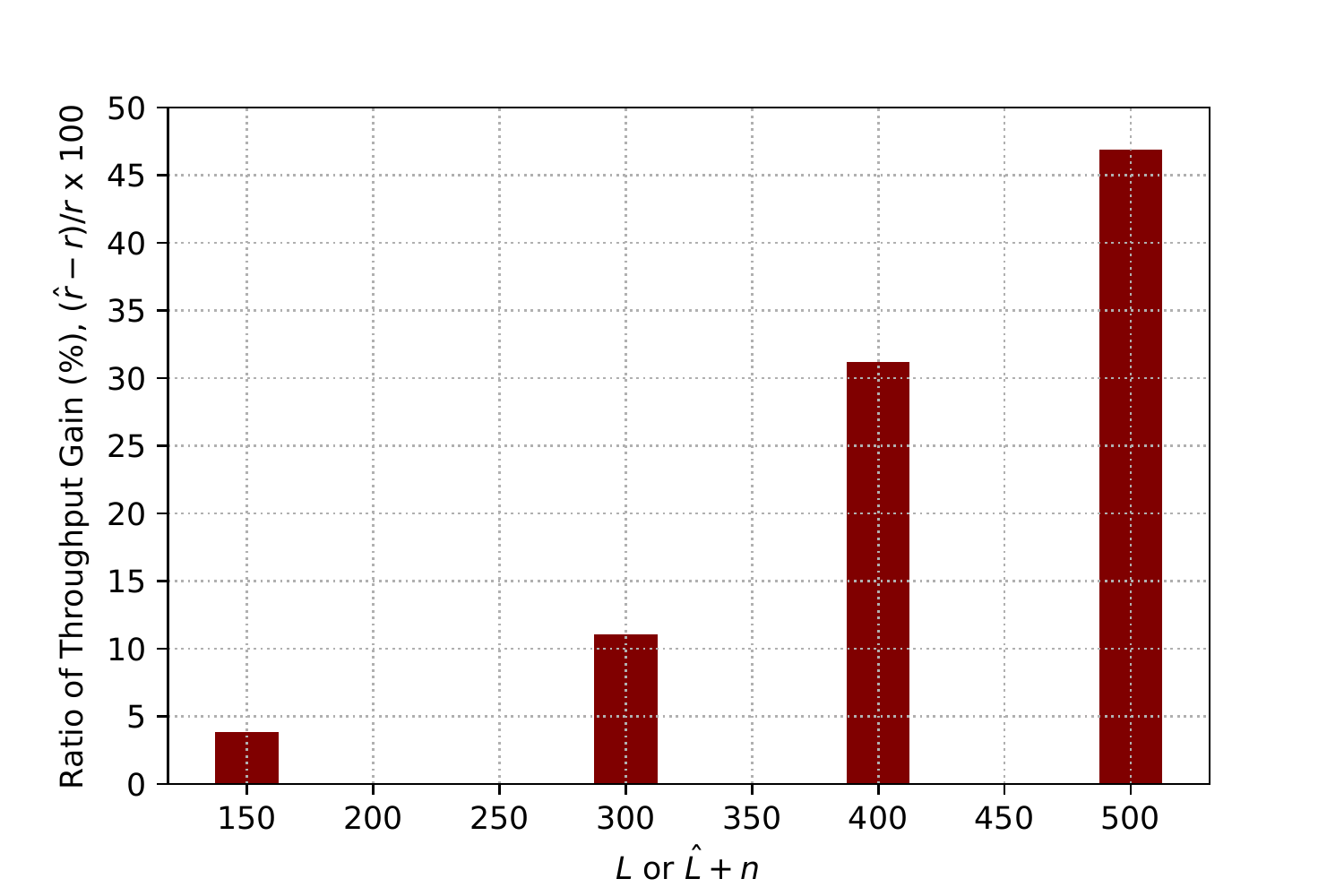}
	\caption{The ratio of throughout improvement for Rayleigh channel.}
	\label{fig:sensing2}
\end{figure}

We show the results for Rayleigh channel and different $\hat L+n$ in Table~\ref{table:summary2}.
Figure~\ref{fig:sensing2} shows the throughput improvement (in terms of $(\hat r -r)/r \times 100$) with and without the GAN. This improvement increases with $L$ (or $\hat L+n$).
With larger $L$ or $\hat L+n$, the throughput is again always decreasing. The misdetection probability tends to decrease also.
If $\hat p_{MD} < 10$\%, the choice of $\hat L=200$ and $n=200$ achieves the best results. The GAN offers more throughput improvement for Rayleigh channels (up to $47\%$)  than AWGN channels (up to $20\%$) because Rayleigh channels are more complex than AWGN channels and thus more (real or synthetic) samples are needed to train a  classifier under Rayleigh channels. 

\begin{table}
	\caption{Comparison of the performance with and without the GAN for Rayleigh channel under different $\hat L$ and $n$.}
	\centering
	{\small
		\begin{tabular}{c|c|c||c|c|c}
		$L$ or & \multicolumn{2}{c||}{No GAN} & \multicolumn{3}{c}{GAN} \\ \cline{2-6}
		$\hat L+n$ & $r$ & $p_{MD}$ (\%) & $n$ & $\hat r$ & $\hat p_{MD}$ (\%) \\ \hline
150 & 4381 & 32.79 & 100 & 4548 & 24.59 \\ \hline
300 & 3646 & 20.40 & 200 & 4050 & 17.32 \\ \hline
400 & 2956 & 7.82 & 200 & 3877 & 8.64 \\ \hline
500 & 2448 & 7.09 & 200 & 3596 & 9.07
		\end{tabular}
	}
	\label{table:summary2}
\end{table}

\section{Conclusion}
\label{sec:conclude}

In this paper, we studied a spectrum coexistence system such as the CBRS band, where the NextG user uses a deep learning-based classifier to identify the incumbent user transmissions by analyzing the spectrum sensing results. A fundamental tradeoff in such a system is that the NextG user needs to spend time to collect spectrum sensing data and build its classifier, which reduces the transmission time and consequently its throughput. One way to improve the performance is using the GAN to generate synthetic sensing data and using such data, along with real sensing data, as the training data to build a classifier. This approach can reduce the sensing time and thus increase the transmission time while keeping the classifier accuracy still high. We showed that for both AWGN and Rayleigh channels, the GAN can achieve a larger NextG user throughput than the approach without the GAN and still protect the incumbent user transmissions. We also showed that the GAN can improve the NextG user throughput more in Rayleigh channels (up to $47\%$) than AWGN channels (up to $20\%$).

\end{document}